\documentclass{article}%
\usepackage{amsfonts}
\usepackage{amsmath}
\usepackage{amssymb}
\usepackage{graphicx}%
\begin{document}

\title{The Flyby Anomaly and the Gravitational-Magnetic Field Induced Frame-Dragging
Effect around the Earth}
\author{Babur M. Mirza\\Department of Mathematics, \\Quaid-i-Azam\ University, {45320 Islamabad}, Pakistan}
\maketitle

\begin{abstract}
The anomalous energy difference observed during the Earth flybys is modelled
here as a dynamical effect resulting from the coupling of the gravitational
and the magnetic fields of the Earth. \ The theoretical analysis shows that
general relativistic frame-dragging can become modified under the Earth's
magnetic field by orders of magnitude. For twelve flyby cases, including the
null results reported in some recent flybys, the predicted velocities
correspond to the observed velocities within the observational error. The
gravito-magnetic effect is also shown to account for the linear distance
relation, time-variation of the anomalous energy, and the reduction in the
anomalous velocity for high altitude flybys near the Earth.\ 

\end{abstract}

\section{Introduction}

Anomalous velocity increase in the flyby orbits has been detected during the
Earth flybys of various spacecraft (Anderson 2008; Anderson 2007; Iorio 2009;
McCulloch 2008; Nottale 2003). The energy change in the reported flybys
corresponds to $10^{-6}$ orders of magnitude on close encounters. The source
of the energy anomaly is not known, and it is now well established that the
effect cannot be due to systematic errors, nor due to dynamical effects, such
as the Earth's rotation, or due to the general relativistic Lense-Thirring
frame-dragging of the spacetime. Studies of various theoretical as well as
phenomenological models has been made by L\"{a}mmerzahl et. al.
(L\"{a}mmerzahl 2018), including the effects due to the atmosphere, ocean
tides, solar wind, and spin--rotation coupling, among others. The general
conclusion of these studies is that these perturbations do not correspond to
the flyby anomaly, where as in phenomenological models it is necessary to pin
down a specific interaction term.

In general the flyby anomaly has various features related to the velocity
increase. It corresponds to a constant acceleration. In some flybys energy
decrease has been observed where as in others the energy increases. However,
these relate to the signs and magnitudes of the spacecraft's initial and final
velocity orientations with respect to the equatorial plane. The empirical
formula of Anderson et.al. (Anderson 2008) gives the energy $K$ corresponding
to the radius $R_{\oplus}$ of the Earth as $K=2\omega_{\oplus}R_{\oplus}/c$,
where $\omega_{\oplus}$ is the angular rotational speed of the Earth, and $c$
is the speed of light. This formula gives the correct order of magnitude for
the orbital energy anomaly, and fits well with the observed change in the
orbital parameters due to the change in the velocity. Further studies by
Anderson et.al. (Anderson 2007) have shown that the energy transfer during
planetary orbital motion occurs as a function of time, which rather than being
a monotonic function of time exhibits local maxima and minima. Some recent
high altitude flybys have detected a null anomalous effect, which is at
variance with the empirical formula $K=2\omega_{\oplus}R_{\oplus}/c$.

In this paper, it is shown that these features, as well as others related to
the flyby anomaly, can be due to the coupling of gravitational and magnetic
fields around the Earth. It has been suggested that since the anomalous energy
change depends on the latitude, Earth's rotation may be generating an effect
much larger than the general relativistic Lense-Thirring frame dragging around
the Earth. We show here that the precise nature of this effect, however, is
not the Lense-Thirring frame-dragging, but a general relativistic coupling of
the magnetic and gravitational fields. Such a coupling can be studied in the
context of Maxwell field equations in the linearized Kerr spacetime, where the
rotation of the source mass induces the dragging of the background spacetime.
In this case the toroidal magnetic field energy density directly modifies the
spacetime frame dragging around a source mass, such as the Earth. The
estimated dimensionless energy $K$ corresponds to the magnetic field and the
radius of the gravitational source, as in the Anderson et. al. formula.
However, the magnetic field intensity reduction with altitude reduces the
magnetic field energy density, hence the dragging effect at large distances
from the Earth. The later case corresponds to the null effect measured by some
recent flybys (Jouannic 2015). Estimate of the anomalous velocity for the
various flybys are shown to be in agreement with the observed velocity change
within the observational errors, as well as the reported null results of the
recent flybys. Throughout we use the standard spacetime coordinates
$(t,r,\theta,\varphi)$.

\section{General Relativistic Toroidal Field Energy-Density}

In a stable equilibrium configuration, the exterior spacetime of a rotating
object of mass $M$ is essentially stationary and axially symmetric. For such a
case the spacetime around the gravitational source can be modelled by the
linearized Kerr metric (Rindler 2006),%
\begin{equation}
g_{\alpha\beta}=\left[
\begin{array}
[c]{cccc}%
-e^{2\Phi(r)} & 0 & 0 & -\omega(r)r^{2}\sin^{2}\theta\\
0 & e^{-2\Phi(r)} & 0 & 0\\
0 & 0 & r^{2} & 0\\
-\omega(r)r^{2}\sin^{2}\theta & 0 & 0 & r^{2}\sin^{2}\theta
\end{array}
\right]  , \tag{1}%
\end{equation}
where
\begin{equation}
e^{2\Phi(r)}=(1-\frac{2M}{r}), \tag{2}%
\end{equation}
$\omega(r)$ is the Lense-Thirring frame-dragging frequency, equal to
$2\mathcal{J}/r^{3}$, and $\mathcal{J}$ is the angular momentum of the
gravitational source (Glendenning 1997). The coupling of the gravitational
field to the magnetic field of such an object can be studied using the general
relativistic form of the Maxwell field equations in curved spacetimes (Landau
\& Lifshitz 1980),
\begin{equation}
\qquad F_{\alpha\beta,\gamma}+F_{\beta\gamma,\alpha}+F_{\gamma\alpha,\beta}=0,
\tag{3}%
\end{equation}%
\begin{equation}
\left(  \sqrt{-g}F^{\alpha\beta}\right)  _{,\beta}=4\pi\sqrt{-g}J^{\alpha},
\tag{4}%
\end{equation}
where $g$ is the determinant of the metric tensor $g_{\alpha\beta}$ given by
equation (1), and $F_{\alpha\beta}$ is the electromagnetic field tensor. It
can be shown (Lichnerowicz 1967) that the assumption of an everywhere finite
current $4$-vector $J^{\alpha}$ leads to the condition that in a co-moving
frame the electric field $4$-vector $E^{\alpha}=0=E_{\alpha}$. Therefore, for
the region exterior to the source, the electromagnetic field tensor has the form,%

\begin{equation}
F_{\alpha\beta}=\sqrt{-g}\epsilon_{\alpha\beta\gamma\delta}u^{\gamma}%
B^{\delta}, \tag{5}%
\end{equation}
and in contravariant form%

\begin{equation}
F^{\alpha\beta}=-\frac{1}{\sqrt{-g}}\epsilon^{\alpha\beta\gamma\delta
}u_{\gamma}B_{\delta}, \tag{6}%
\end{equation}
where $\epsilon_{\alpha\beta\gamma\delta}$ is the four index Levi -Civita
symbol. Here a convenient and physically meaningful choice of a co-moving
observer is the one belonging to the class of observers ZAMOs, having zero
angular momentum. This is an observer circling the gravitational source with
angular speed $\omega$ at a fixed radial distance $r$, per unit length, and
polar angle $\theta$, being brought into motion by dragging of the background
spacetime (Rindler 2006). By definition, for a ZAMO $u_{r}$ and $u_{\theta
\text{ }}$ vanish, and the velocity four vector compatible with the metric is
given by%
\begin{equation}
u^{\alpha}=(e^{-\Phi(r)},0,0,\omega e^{-\Phi(r)}),\quad u_{\alpha}%
=(-e^{\Phi(r)},0,0,0). \tag{7}%
\end{equation}

\noindent\emph{\ }Given the current $4$-vector $J^{\alpha}$, the metric tensor
$g_{\alpha\beta}$, and the velocity $4$-vector $u^{\alpha}$, the set of
equations (3) and (4) can be solved to give the magnetic field as a function
of $(t,r,\theta,\varphi)$. The full set of Maxwell equations (3) and (4) gives
a system of eight coupled partial differential equations. Solutions to the
Maxwell equations in axially symmetric spacetimes has been found (Mirza 2017;
P\'{e}tri 2013; Mirza 2007; Herrera et al. 2006; Oron 2002) which satisfy the
asymptotic boundary conditions of finite energy density, and other physical
conditions such as a bounded magnetic field outside the event horizon. Using
equations (5) and (6) in equations (3) and (4), it is easily seen that for
$\omega\neq0$ and $B_{t}\neq0$, the magnetic field around the star has the
form,%
\begin{equation}
(B^{\alpha})=B_{0}\left(  0,0,A(\theta)\sin\theta\sin\xi,\cos\xi\right)
/(u_{t}r^{2}\sin^{2}\theta), \tag{8}%
\end{equation}
where $B_{0}$ is the magnetic field intensity constant, $\xi=\varphi-\omega t$
is the travelling wave variable, and $A(\theta)=\int\left(  1/\sin
\theta\right)  d\theta$. Notice that the magnetic field (8) is due to the
frame-dragging effect of the background spacetime which vanishes, otherwise,
for $\omega=0$. For a free-falling observer, such as the ZAMO, $\omega
=d\varphi/dt$, implies that $\xi=\xi_{0}=0$, and the non-vanishing component
of the magnetic field is the toroidal component. This is also because the
integral $A(\theta)=\int\left(  1/\sin\theta\right)  d\theta$ is equal to zero
for a closed loop around the source. The dragging effect is maximum in the
equatorial plane $\theta=\pi/2$ of the gravitational source, therefore using
$B_{T}=\sqrt{B^{\alpha}B_{\alpha}}$, the magnetic field component effective in
the equatorial plane of the Earth is the toroidal field component,
\begin{equation}
B_{T}=-\frac{B_{0}}{r\sqrt{1-\frac{R_{s}}{r}}}\cos(\varphi-\omega t),\text{
}r>R_{s}, \tag{9}%
\end{equation}
where $R_{s}=2GM/c^{2}$ is the Schwarzschild radius for the source mass $M$.
Also from equations (2) and (7), we have $u_{t}=-\sqrt{1-R_{s}/r}$. For a
slowly rotating gravitational source like the Earth $\omega<<1$, and the
cosine factor depends on the choice of the coordinate $\varphi$. For the
maximum dragging effect we take $\varphi=0$. The radial drag therefore
corresponds to $B_{0}/\left(  r\sqrt{1-R_{s}/r}\right)  $ in the equatorial
plane of the rotating gravitational object.

The total energy density in the toroidal field is given by (Landau \& Lifshitz
1971),
\begin{equation}
\epsilon_{B}=\int_{0}^{R}4\pi r^{2}B_{T}^{2}dr, \tag{10}%
\end{equation}
where $R$ is the radius of the source mass. To calculate the energy density
(10), we notice that the toroidal field has a discontinuity at the
Schwarzschild radius $r=R_{s}$. We thus divide the region of integration in
equation (10) as the interior region $r\leq R_{s}$ and the exterior region
$r>R_{s}$, such that,%
\begin{equation}
\epsilon_{B}=\int_{0}^{R_{s}}4\pi r^{2}B_{T}^{2}dr+\int_{R_{s}}^{R}4\pi
r^{2}B_{T}^{2}dr. \tag{11}%
\end{equation}
The first integral can be calculated using $B_{T}=B_{0}/r\sqrt{1-R_{s}/r}$, we
therefore obtain,
\begin{equation}
\int_{0}^{R_{s}}4\pi r^{2}B_{T}^{2}dr=4\pi B_{0}(R_{s}+R_{s}Lt_{\Delta
r\rightarrow0}\ln\Delta r-R_{s}\ln\mid R_{s}\mid), \tag{12}%
\end{equation}
where $\Delta r=\mid r-R_{s}\mid$. Similarly, for the second integral, we
have,
\begin{equation}
\int_{R_{s}}^{R}4\pi r^{2}B_{T}^{2}dr=4\pi B_{0}(R-R_{s}+R_{s}\ln\mid
R-R_{s}\mid-R_{s}Lt_{\Delta r\rightarrow0}\ln\Delta r). \tag{13}%
\end{equation}
Putting from (12) and (13) into equation (11), we have for for the exterior
region $R>R_{s}$,
\begin{equation}
\epsilon_{B}=4\pi B_{0}^{2}\left[  R+R_{s}\ln\left(  R-R_{s}\right)  \right]
. \tag{14}%
\end{equation}

This energy density, although due to the coupling of the magnetic to the
gravitational field, contributes to the total energy density around the source
mass. Notice that the gravitational field energy depends on the Schwrazschild
factor $R_{s}=2GM/c^{2}$. Since for the Earth, $R>>R_{s}$, therefore the
second term in equation (14) can be dropped, and the formula for the
dimensionless anomalous energy per unit energy density is,%
\begin{equation}
K=4\pi B_{0}^{2}R,\text{ } \tag{15}%
\end{equation}
for $R>R_{s}$.

Some key features of formula (15) are noticeable. (1) The anomalous energy is
linearly dependent on the radius of the Earth, as in Anderson et. al. (2008).
We thus identify $2\omega/c$ as the magnetic energy density $4\pi B_{0}^{2}$,
which is also dimensionally consistent by equation (15). (2) When the test
body moves towards the source, the total energy density will be reduced, since
the work done is opposite to the force field, where as contrary will be the
case for an outward motion. (3) For a time-varying field, the anomalous energy
should exhibit an oscillatory trend with time as $\cos^{2}(\varphi-\omega t)$,
as reported in Anderson et.al. (2007).

The energy given by equation (15) is the kinetic energy per unit normalized
energy density $E$ in the equatorial plane of the Earth. The effective
components along the velocity vectors in the plane of the orbit are $4\pi
B_{0}^{2}R\cos\phi_{i}$ and $4\pi B_{0}^{2}R\cos\phi_{o}$, where $\phi_{i}$
and $\phi_{o}$ are the incoming and outgoing orbital latitudes of the flyby,
respectively. Also, since for the velocity change $dV/V_{\infty}%
=dE/2E=K(\cos\phi_{i}-\cos\phi_{o})$, we have,%
\begin{equation}
dV=4\pi B_{0}^{2}RV_{\infty}(\cos\phi_{i}-\cos\phi_{o}) \tag{16}%
\end{equation}
where $V_{\infty}$ is the asymptotic (hyperbolic) excess velocity of the
flyby. The net velocity change is twice the velocity increase (or decrease)
along the incoming or outgoing trajectories. The average surface magnetic
field of the Earth approximately varies from $0.3G$ to $0.6G$. We take the
surface magnetic field in calculations (Table 1) to be $0.5G$ in cgs units.
The corresponding magnetic field in SI units is of the order $0.0126\times
0.5A$ per unit distance. According to equation (15), the magnetic field
weakens as $1/\sqrt{h}$ where $h$ is the height of the test particle above the
Earth's surface. For flyby encounters the average distance of $10^{3}km$
implies a reduction in the magnetic field by an order $10^{-3}$.

In Table 1, the estimated velocity change $dV_{i}$, using equation (16), is
compared with the observed velocity change $dV_{o}$ for flybys of twelve
spacecraft. In the case of last six flybys the average asymptotic speed is
taken to be the average speed of $10km/s$. The minimum altitude is taken to be
the height $h$ above the Earth surface, measured during close encounters. In
relation (15) the intrinsic magnetic field $B_{0}$ and the radius $R$ of the
gravitating body are the only input parameters needed to estimate the
anomalous energy. For the Earth, the estimated dimensionless anomalous energy
$K$ comes out to be about $3.1\times10^{-6}$, which is comparable with the
observed average energy difference for the flybys. For planets with intrinsic
dipolar magnetic fields of similar orders of magnitude, but larger radii, the
estimated anomalous energy near the surface also has a comparable magnitude
for $K$.%

\begin{tabular}
[c]{|l|l|l|l|l|l|l|l|}\hline
Mission & $V_{\infty}$ ($km/s$) & $h$ ($km$) & $\phi_{i}$ & $\phi_{o}$ &
$\Delta$ & $dV_{o}$ ($mm/s)$ & $dV_{p}$ ($mm/s$)\\\hline
Galileo-I & $8.949$ & $960$ & $-12.52$ & $-34.15$ & $0.149$ & $3.92\pm0.3$ &
$4.133$\\\hline
Galileo-II & $8.877$ & $303$ & $-34.26$ & $-4.87$ & $-0.17$ & $-4.6\pm1.0$ &
$-4.678$\\\hline
NEAR & $6.851$ & $539$ & $-20.76$ & $-71.96$ & $0.626$ & $13.46\pm0.01$ &
$13.295$\\\hline
Cassini & $16.010$ & $1175$ & $-12.92$ & $-4.99$ & $-0.022$ & $-2\pm1$ &
$-1.09$\\\hline
Rosetta & $3.863$ & $1955$ & $-2.81$ & $-34.29$ & $0.172$ & $1.80\pm0.03$ &
$2.058$\\\hline
Messenger & $4.056$ & $2347$ & $31.44$ & $-31.92$ & $5\times10^{-3}$ &
$0.02\pm0.01$ & $0.062$\\\hline
Rosetta-II & $10.0$ & $5301$ & $-10.80$ & $18.60$ & $0.034$ & $0$ &
$0.1$\\\hline
EPOXY-I & $10.0$ & $15614$ & $-5.18$ & $16.90$ & $0.039$ & $0$ &
$0.12$\\\hline
EPOXY-II & $10.0$ & $43415$ & $17.03$ & $64.12$ & $0.519$ & $0$ &
$1.6$\\\hline
Rosetta-III & $10.0$ & $2480$ & $18.49$ & $24.37$ & $0.040$ & $0$ &
$0.12$\\\hline
EPOXY-III & $10.0$ & $30404$ & $-63.88$ & $-8.04$ & $0.549$ & $0$ &
$1.7$\\\hline
Juno & $10.0$ & $561$ & $14.17$ & $39.50$ & $0.197$ & $0$ & $0.6$\\\hline
\end{tabular}

Table I: The anomalous change in the flyby velocity as predicted ($dV_{p}$)
from equation (16), and compared with the observed ($dV_{o}$) velocity change
for flyby missions during 1990-2013. The asymptotic velocity $V_{\infty}$ is
measured in $km/s$ (Anderson et. al. 2008; Jouannic et. al. 2015), and
$\Delta=\cos\phi_{i}-\cos\phi_{o}$. The scaled dimensionless radius
$R_{\oplus}$ for the Earth is in per unit length of $km$, where as the average
surface magnetic field of the Earth is $0.5G$ over unit distance per $km$.

\section{Conclusions and Discussion}

In Table 1, data (Anderson et.al. 2008; Jouannic et. al. 2015) of six flyby
missions, including the observed and calculated velocity changes, is given.
The anomalous velocity $dV$ exhibits positive and negative signs, and in the
case of Messenger flyby the anomalous effect is close to zero. We notice that
the dragging effect is maximum in the equatorial plane of the Earth and
decreases towards the poles. Resultingly, for a small ingoing declination
angle $\phi_{i}$ the dragging effect is larger as compared to a larger
outgoing declination angle. The velocity change is therefore positive as in
the case of Galileo-I, NEAR, and Rosetta flybys. Also, since the velocity
change depends on the flyby asymptotic velocity $V_{\infty}$ relative to the
Earth, the spin rotation of the Earth determines the sign of the anomalous
velocity change. For the case of Messenger the ingoing and outcoming
declination angles are approximately equal, hence the drag along the orbital
motion towards and away from the source, approximately cancels. In this case
the net velocity change is small.

We recover the Anderson et.al. empirical formula $K=2\omega_{\oplus}R_{\oplus
}/c$ by observing that near the Earth's surface change in the energy density
due to the spacetime dragging is proportional to the change in the rotational
velocity of a particle placed close to the surface of the Earth. Therefore, in
one cycle, $\omega$ varies as $2\pi B_{0}^{2}$. With speed of light $c$ being
the constant of proportionality, this yields the relation $K=2\omega R/c$.
However, in the empirical formula of Anderson et. al. the effects of rotation
do not depend on the flyby's distance from the Earth. We observe that \ for
the dimensionless energy $K$, equation (16) implies that $B_{0}$ is reduced as
$1/\sqrt{h}$, where $h$ is the height of the test particle above the Earth's
surface. The magnetic field intensity thus decreases by a factor $10^{-3}$
units at an altitude of $10^{6}m$, which is the approximate distance of the
flybys in Table 1. In the high altitude flyby encounters, such as in the case
of Rosetta-II, EPOXY-II, and EPOXY-III, the average distance of the flyby in
trajectory is approximately $10-50$ times the flyby altitudes reported in
Anderson et.al. The resulting anomalous velocity change therefore diminishes
as $B_{0}^{2}\sim10^{-1}-2\times10^{-2}$ units, hence tends to lie below the
range of the observational error in high altitude flybys. As reported in
Jouannic et.al., no anomalous velocity change was observed for Rosetta-II, and
for EPOXY-II and III flybys. In closed orbital motion, however, the energy
density have the cumulative effect over time. Thus for the lunar orbit, net
velocity drag along the radial direction is $\sqrt{4\pi R}B_{0}\approx
10^{-9}m/s$ per year. This is comparable with the observed lunar orbital
recession speed (Anderson \& Nieta 2009).

The coupling of gravitational field to the magnetic field in extreme
gravito-magnetic environments is well-studied, and leads to an enhancement of
the magnetic field energy density of a star. This however contributes to the
total energy density of the field around a massive object, possessing both
magnetic field and rotation. In neutron stars and magnatars the wrapping of
the spacetime around the star can enhance the energy density up to several
orders of magnitude (Mirza 2017). Analogous conditions exist in almost all
gravitationally bound systems with a dipolar magnetic field and a toroidal
field component. Although, in such cases, the magnetic field is weak and
rotation is slow, the coupling effect can possess measurable magnitude. The
effect derived here is due to rotation but cannot strictly be the
Lense-Thiring frame-dragging, where only the gravitational field is involved,
and dragging occurs close to the surface of the star. Magnetic field causes a
more effective dragging of the spacetime, thus enhances the spacetime
frame-dragging up to comparatively larger distances.

\bigskip

\begin{thebibliography}{99}                                                                                               %


\bibitem {[1]}Anderson, J. D., Campbell, J. K., and Nieto, M. M., 2007, New
Astron. Rev. 12, 383.

\bibitem {[2]}Anderson, J. D., Campbell, J. K., Ekelund, J. E., Ellis, J., and
Jordan, J. F., 2008, PRL 100, 091102.

\bibitem {[3]}Anderson, J. D., and Nieto, M. M., 2007, arXiv:0907.2469.

\bibitem {[4]}Antreasian, P. G., Guinn, J. R., 1998, Astrodynamics Specialist
Conference (AIAA, Washington), pp. 98-4287.

\bibitem {[5]}Glendenning, N.K., 1997, Compact Stars, (Springer-Verlag, New
York), pp. 254-255.

\bibitem {[6]}Herrera, L., Gonz\'{a}lez, G. A., Pach\'{o}n, L. A., \& Rueda,
J. A\textit{.,} 2006, Class. Quantum Grav. 23, 2395.

\bibitem {[7]}Iorio, L., 2009, Scholarly Research Exchange 2009.

\bibitem {[8]}Jouannic, B., Noomen, R., and van den IJssel, 2015, JAA.
Proceedings of the international symposium on space flight dynamics, ed. R.
Kahle (DLR, Munich), pp. 1-17.

\bibitem {[9]}L\"{a}mmerzahl, C., Preuss, O., \& Dittus, H., 2018, arXiv:gr-qc/0604052.

\bibitem {[10]}Landau, L. D. \& E.M. Lifshitz, E. M.,1971, The Classical
Theory of Fields, (Pergamon, New York), p. 67-74.

\bibitem {[11]}McCulloch, M. E., 2008, MNRAS Letters, 389, L57.

\bibitem {[12]}Mirza, B.M., 2007, Int. J. Mod. Phys. D\textbf{16} 1705 (2007).

\bibitem {[13]}Mirza, B.M. 2017, ApJ. 847, 73.

\bibitem {[14]}Nottale, L., 2003, arXiv preprint gr-qc/0307042.

\bibitem {[15]}Oron, A., 2002, Physical Review D66, 023006.

\bibitem {[16]}P\'{e}tri, J., 2013, MNRAS 433, 986.

\bibitem {[17]}Rindler, W., 2006, Relativity: Special, General, and
Cosmological, second edition, (Oxford University Press, Oxford), pp.337-339.
\end{thebibliography}
\end{document}